# 10 Inventions on laptop keyboards
# -A study based on US patents

**Umakant Mishra**
Bangalore, India
umakant@trizsite.tk
http://umakant.trizsite.tk

**Contents**

## 1. Introduction

A desktop keyboard has several sections like character key section, navigation key section, numeric key section, and function key section etc. each consisting of several number of keys. However, a laptop computer does not have so much of space to accommodate all these keys into the keyboard. There are several considerations while designing a laptop keyboard.

### Necessary features

- The laptop keyboard should accommodate as many keys of a full size keyboard as possible within the limited space of the keyboard.
- The size of the keyboard should be small to fit inside the laptop box.

### Desirable features

- It should be light to carry.
- It may be expandable for comfort of operation.
- It may have additional special keys.
- It should be tiltable, adjustable and ergonomic.
- It should be collapsible to consume less storage.

# 2. Inventions on Laptop keyboards

## 2.1 A tiltable keyboard for Portable computers (Patent 5168427)

### Background problem

The keyboards on portable computers are flat and parallel to its base. When the laptop is on a table, the human arm is at an angle to the keyboard and not comfortable to operate a flat keyboard. For example when the laptop is on the table, a downwardly sloping orientation would make it convenient to operate.
Thus it is beneficial to have the ability of tilting the keyboard according the position of the laptop to the user.

### Solution provided by the invention

Clancy et al. provided a solution to solve the above problem (Patent 5168427, assigned to Compaq Computer Corporation, Issued in Dec 1992). According to the invention, the keyboard structure is recessed within an open topped base housing portion of a compact notebook or laptop computer.
Front corner portions of the keyboard structure are secured to the base housing in a manner permitting the keyboard structure to be pivoted relative to the base housing.

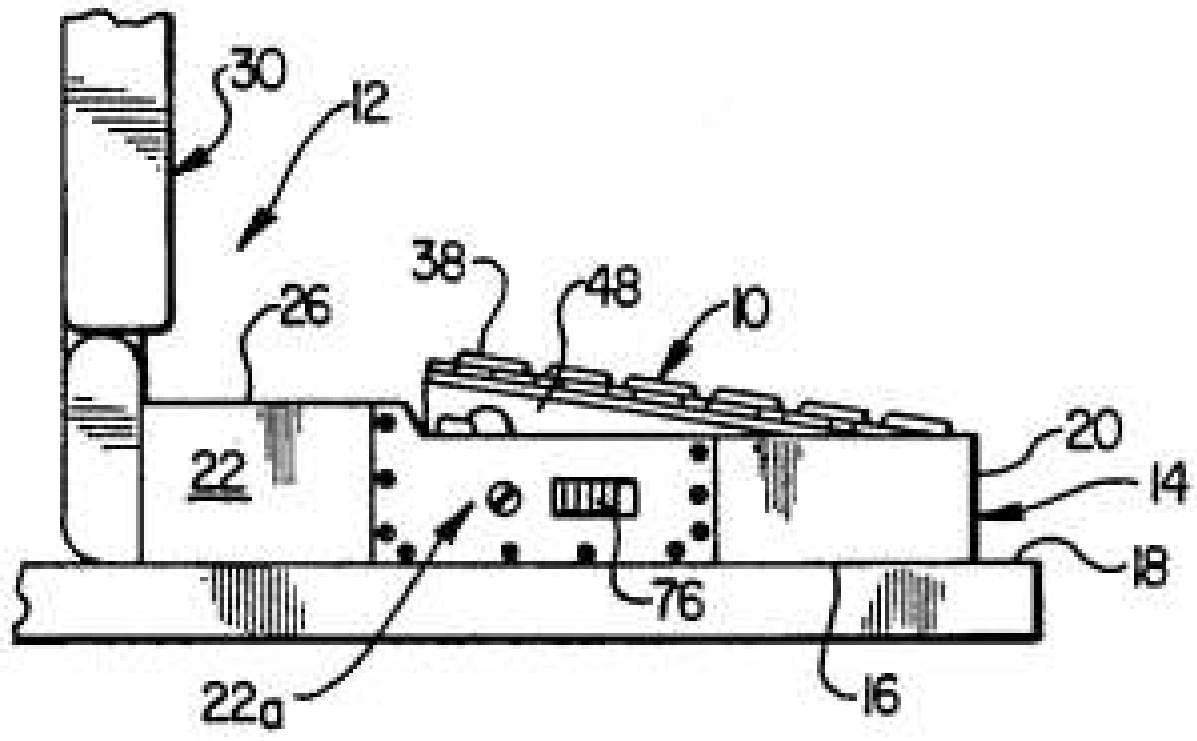

During storage or transport, the keyboard is locked with the housing in a parallel condition to the base housing. When the laptop is used, the keyboard lock can be opened to pop up the keyboard to an ergonomically improved tilted orientation.

The structure of the keyboard slopes forwardly and downwardly for comfortable use. Although the keyboard can still be used in a locked flat position.

### TRIZ based analysis

The keyboard should be at the same angle as the user's arm for comfortable operation **(IFR)**.

We want the keyboard of the laptop with a forwardly and downwardly sloping orientation, but we don't want to increase the overall exterior dimension of the laptop **(Contradiction)**.

The invention uses a tiltable keyboard, which can be sloppy while in use and flat when closed inside cover of the laptop **(Principle-15: Dynamize)**.

## 2.2 Folding and sliding keyboard for Personal Computer (Patent 5267127)

### Background problem

The portable computers or laptops need small size keyboards that should fit into the laptop box. A typical problem in reducing the size of a keyboard is its inconvenience of operating compared to a standard sized keyboard. It is necessary to have a large keyboard while typing but a small keyboard while not using.

### Solution provided by the invention

Pollitt invented a folding and sliding keyboard (Patent 5267127, assigned to IBM, Issued in Nov 93). The size of the keyboard is same as a standard keyboard when used, but very small (almost half) when dissociated in a stored position. The keyboard in the invention is segmented to two halves. The portions of the keyboard are designed for a pivotal movement on a keyboard axis.

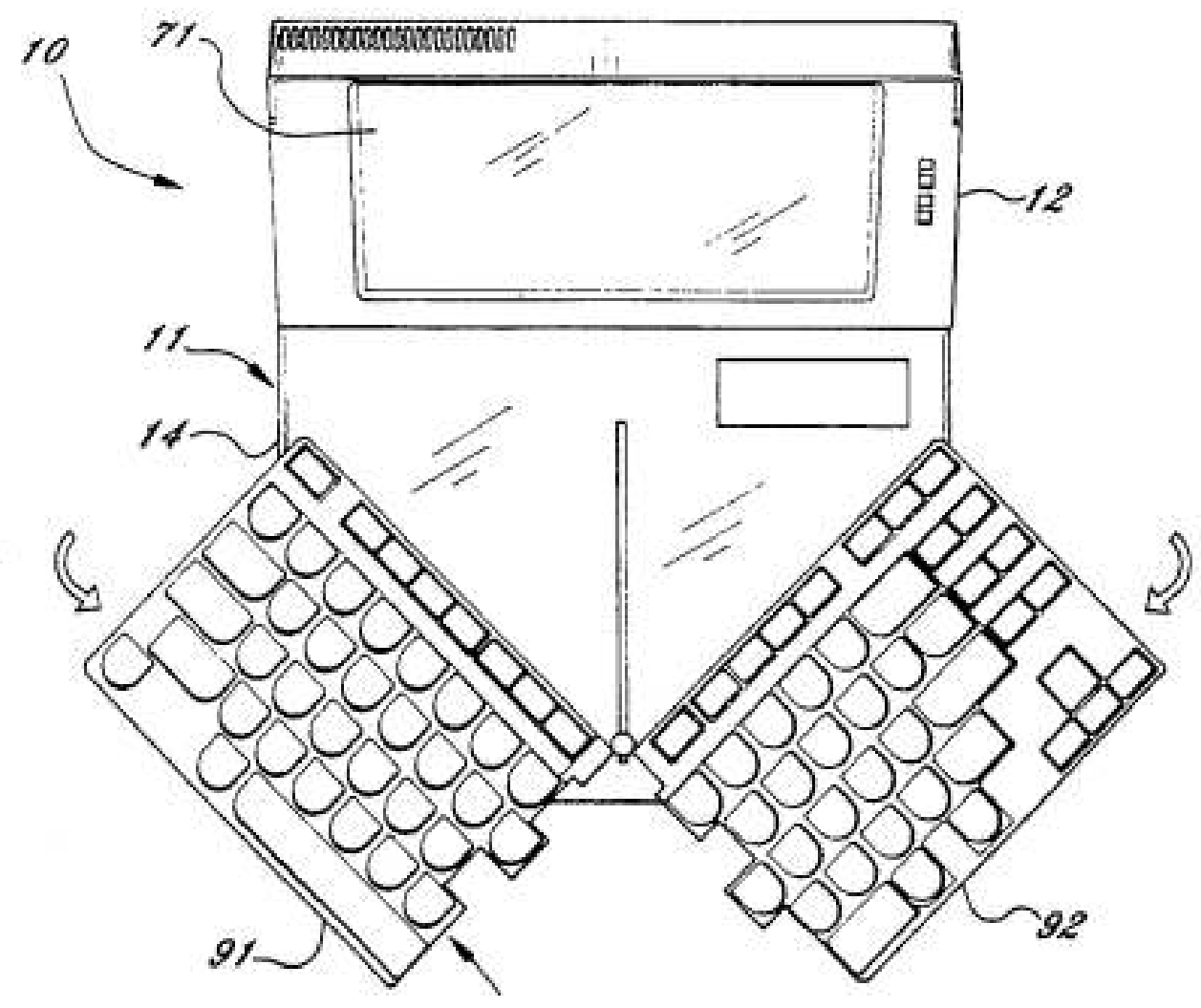

When the keyboard halves are slided to join together they form a complete keyboard of a standard size. When they are stored, they are slided back to have a dimension to fit within the enclosure.

### TRIZ based analysis

The keyboard should change its size to big or small as and when required **(Ideal Final Result)**. The size of a keyboard should be that of a standard desktop keyboard for typing comfort. But the size of the keyboard has to be small to fit into the laptop box **(Contradiction)**.

The invention segments the keyboard into two halves **(Principle-1: Segmentation)**.

The keyboard portions are designed for a pivotal movement, which facilitates folding and sliding **(Principle-15: Dynamize)**.

## 2.3 Ergonomic Folding keyboard for laptops (Patent 5410333)

### Background problem

The small size keyboards used in laptops and palmtops are not convenient for typing. These keyboards reduce efficiency and cause Repetitive Stress Injuries.

### Solution provided by the invention

Kevin Conway disclosed a method of ergonomic computer keyboard (Patent 5410333, Issued in Apr 95), which solves this classical contradiction on the size of the keyboard. The invented keyboard includes two keyboard portions having keyboard keys disposed on both the portions. The keyboard portions are interconnected to give the shape of a full keyboard.

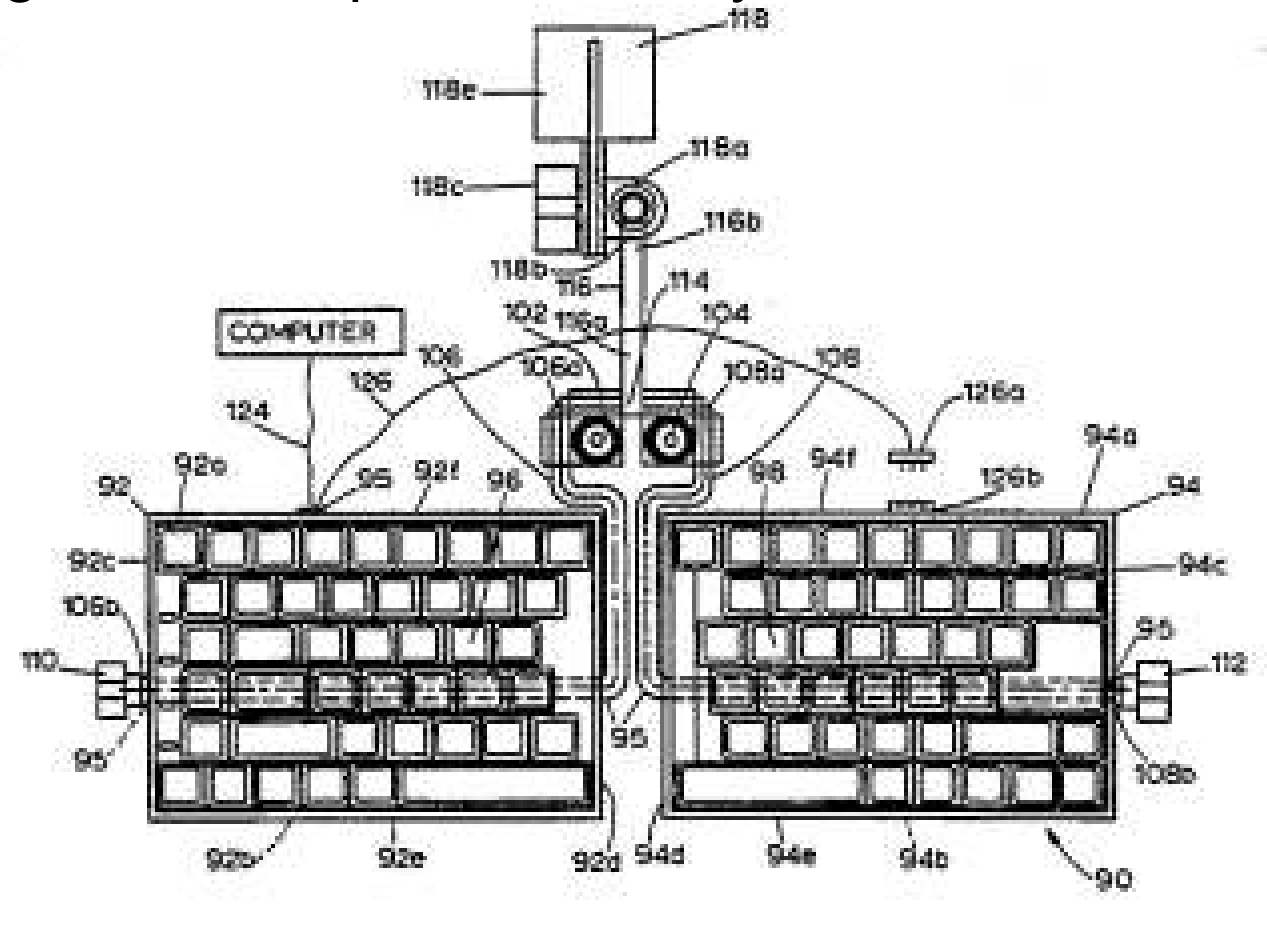

### TRIZ based analysis

The keyboard should be large on use and small on storage and carriage. A reduction in the size of the keyboard reduces its efficiency. An increase in the size of the keyboard prevents its storage and carriage **(Contradiction)**.

The invention splits the keyboard into two **(principle-1: Segmentation)**, opens the keyboard when use and folds the keyboard when stored **(Principle-15: Dynamize)**.

## 2.4 Tri-fold personal computer with touchpad and keyboard (Patent 5926364)

### Background

The notebook computers are small, light but expensive. The desktop computers are powerful, cheaper but large. There is a need to combine the elements of both desktop and a notebook computer and get the benefits of both.

### Solution provided by the invention

Karidis disclosed a solution (patent 5926364, assigned to IBM, July 99) of a hybrid packaging design for a portable personal computer. The invention comprises a tri-fold mechanical structure with a touch-screen display screen and a detachable keyboard, which is easy to pack with the computer. It has a stylus, a wireless remote control and other devices as well. The user can operate through the touchpad using fingers or a stylus.

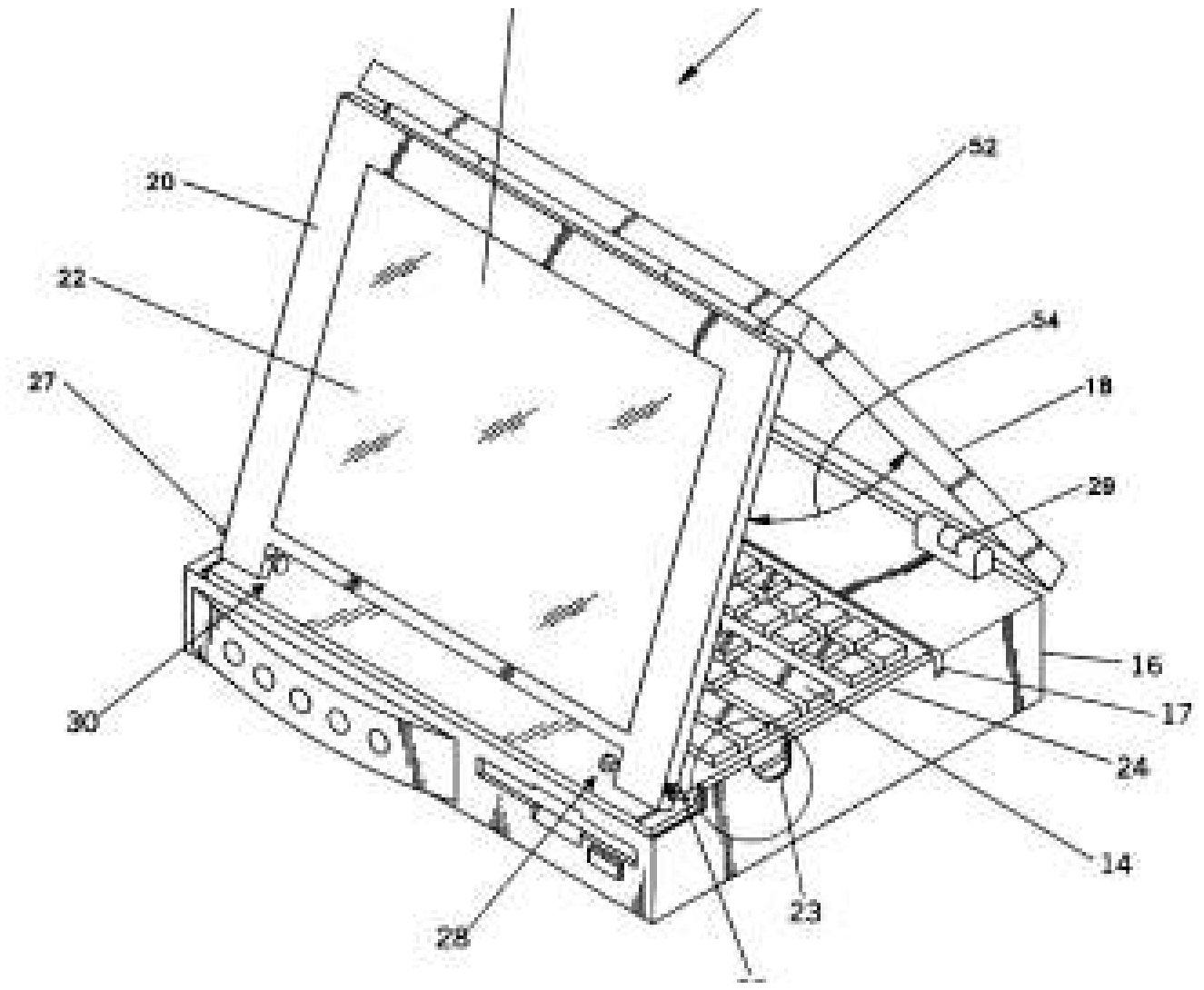

### TRIZ based analysis

The small notebook computer should contain all the features of a desktop computer **(desired result)**.

Normally a notebook has a folded display unit **(Principle-15: Dynamize)**. The invention proposes to increase the number of folds to threefold **(Principle-17: Another dimension)**.

It uses a detachable keyboard **(Principle-15: Dynamize)**.

## 2.5 Expandable keyboard for small computers (Patent 5938353)

### Background problem

The small computers need small keyboards for carrying along with the computer. The user reduces his typing speed when the size of the keyboard is even slightly reduced.

One option is make a folding structure of the keyboard. The folding structure of the keyboard, although takes less surface space, increases the depth (or height) significantly. There is a need to reduce the size of the keyboard without increasing its height unlike a folding keyboard.

### Solution provided by the invention

Butler invented an expandable keyboard (Patent 5938353, Aug 99) for small computers whose keys have interlocking side edges and are mounted on a scissor-linkage that allows said keyboard to contract to a width that is much narrower than the width of a standard desktop computer keyboard.

The invention is based on the truth that a significant amount of space exists between the raised side edges of any pair of adjacent keys in a standard (19 mm) pitch. This gap between the keys is about half of the width of the key. The problem can be addressed if these gaps are available or expanded in “open” position, yet eliminated or contracted in “closed” position. The elimination of gaps between the keys can reduce a 28-29 cm keyboard to 17-18 cm length without increasing the depth.

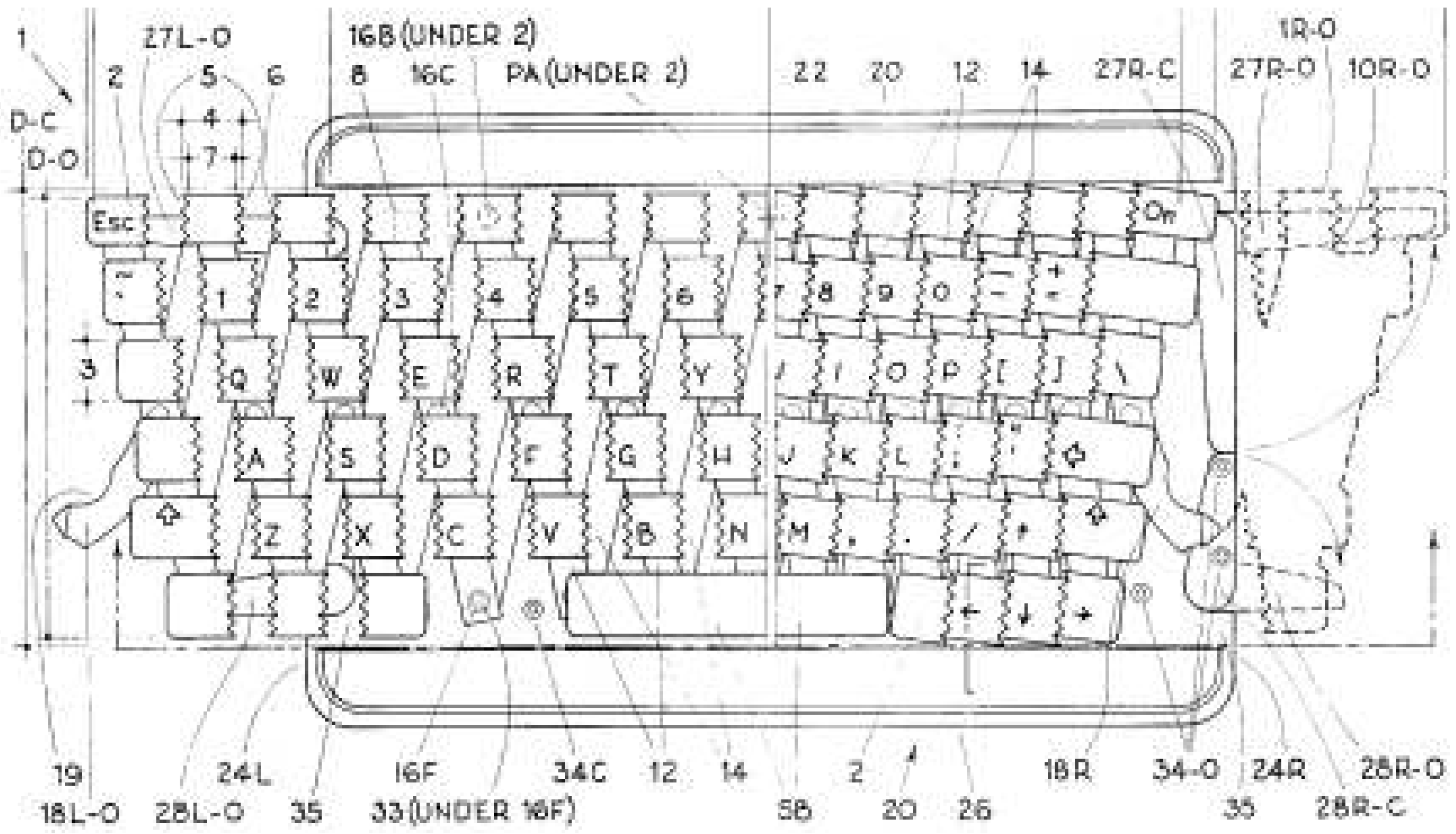

### TRIZ based analysis

The keyboard should be small to carry and big to provide typing comfort **(Contradiction)**.

The invention eliminates the gap between keys when the keyboard is packed for carrying and expands the gap between keys when used for typing **(Principle-34: discard and recover)**.

## 2.6 Computer bottom keyboard incorporating arrangement for enhanced cooling (Patent 5978215)

### Background problem

The portable computers are compact in size and, unlike desktop PCs, have no space for ventilation. This makes the notebook get heated quickly. One of the methods to reduce heat is using high power cooling fans. But unfortunately they require more battery consumption. There is need to increase the cooling capacity of the portable computers.

### Solution provided by the invention

Chiu et al. disclosed a keyboard for portable computers having enhanced cooling mechanism (patent 5978215, assigned to IBM, Nov 99). According to the invention the portable computer has outwardly slidable compartments above the keyboard. When these compartments are slided out, the structure provides increased surface areas for dissipation of heat.

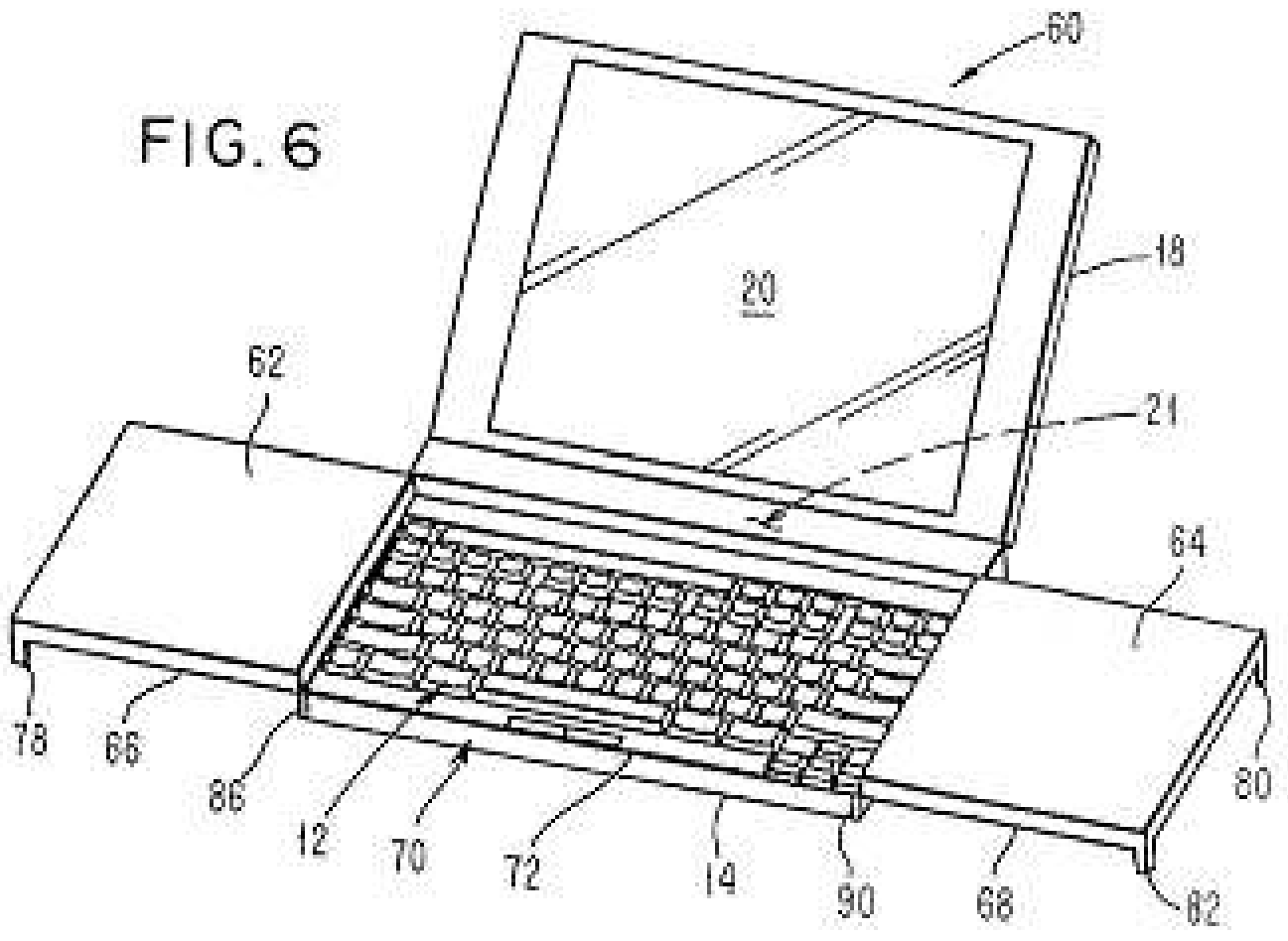

### TRIZ based analysis

The portable computer should automatically cool by itself **(Ideal Final Result)**.

The invention exposes more internal components of the computer for faster dissipation of heat **(Principle-35: Parameter change)**.

One of methods used by this invention is an openable or slidable laptop body **(Principle-15: Dynamize)**.

## 2.7 Keyboard and notebook type computer (5995024)

### Background problem

The operating surface of the laptop keyboard is very small because of the miniaturization of the computer. The low operating surface of the laptop keyboards makes them inconvenient for typing. How to increase the operating surface without increasing the size of the keyboard?

### Solution provided by the invention

Kambayashi et al. invented a keyboard (Patent 5995024, Assigned to Fujitsu Limited, Nov 99) that uses a similar method to provide more operating space. The invention realizes that the transverse directional length is useful for typing comfort; hence we can reduce the vertical length of the keyboard without affecting the typing comfort.

According to the invention the size of the keys are reduced in the vertical direction and gaps between keys are reduced in the vertical direction without changing the size of the keys or gaps between the keys in the transverse direction. This makes the length of the keyboard larger in transverse direction than that in the vertical direction.

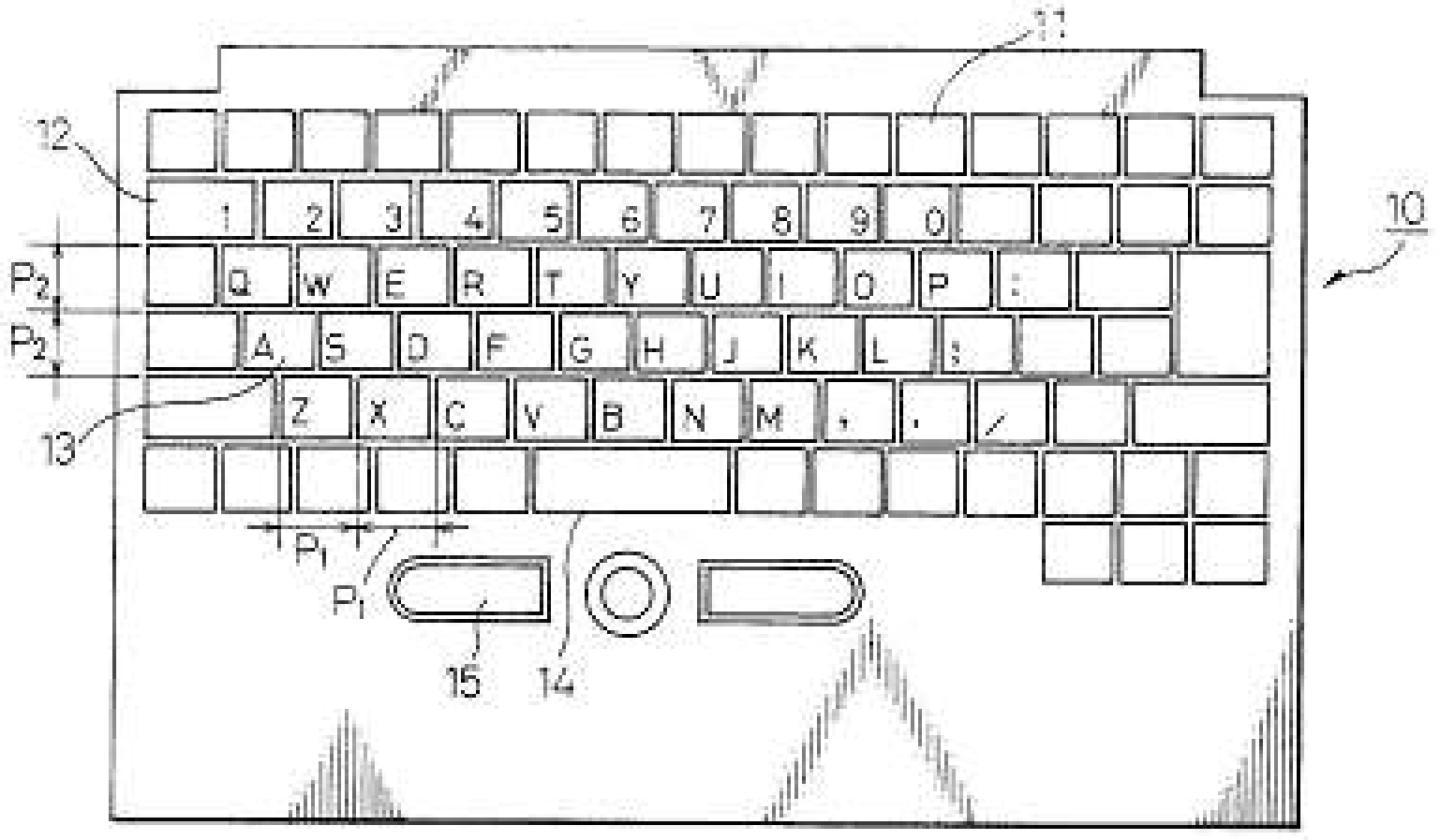

As the vertical directional length of the upper surface of the keyboard is shortened, a relatively large armrest area is obtained, improving the operability results. As the transverse directional length of the keytop is longer than the vertical directional length, it provides the operational comfort and reliability.

### TRIZ based analysis

the keyboard should provide larger operating space for typing comfort, but providing more operating space will eventually increase the size of the keyboard that will not fit into a laptop. **(Contradiction)**

The invention reduces the vertical size of the keys and vertical space between keys, which reduces the vertical size of the keyboard keeping the transverse size in tact. This provides operating space as well as more armrest area **(Principle-2: Taking out, Principle-17: Another dimension)**.

## 2.8 Keyboard having an integral heat pipe (Patent 6215657)

### Background problem

The processor of the computer generates a lot of heat during operation. It's necessary to transfer the heat from the IC for smooth operation of the computer especially for portable computers, which does not provide enough ventilation for heat elimination.

There are various alternative solutions used to solve this problem. (i) Use a metal plate as the heat transfer media, (ii) use the keyboard support plate as the heat transfer media, (iii) heat sinks having integral fans etc. Each having some advantages and disadvantages.

### Solution provided by the invention

Rakesh Bhatia invented a keyboard having an integral heat pipe (Patent 6215657, assigned to Intel corporation, Apr 01) for heat transmission. The heat pipe provides structural support to the keyboard and is thermally coupled to one or more heat generating components within the computer system. The heat pipe replaces the prior art keyboard support plates. The heat pipe comprises several heat pipes which are arranged in a parallel configuration. Each heat pipe is divided by sidewalls and is sealed to contain a two-phase vaporizable liquid, which serves as the working fluid for the heat pipe.

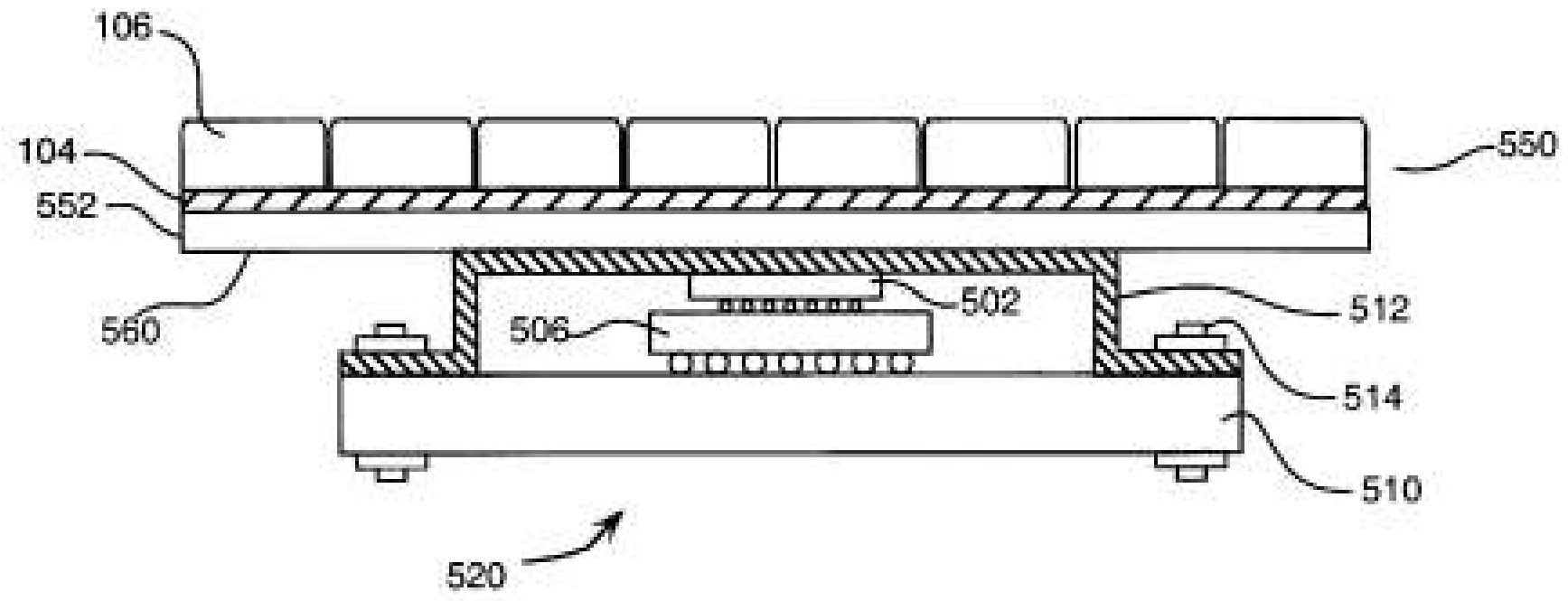

### TRIZ based solution

The invention uses heat pipes to transfer heat from inside **(Principle-31: Hole)**.

## 2.9 Portable computer with ergonomic keyboard (Patent 6304431)

### Background problem

Conventionally the keyboard is statically attached to a portable computer. This needs the user to position himself straight to operate the computer which may not be hygienic or comfortable.  There are some ergonomic keyboards for the desktops but nothing for the laptops. There is a need for an ergonomic keyboard for laptop.

## Solution provided by the invention

Myung-Jung Kim invented an ergonomic keyboard (Patent Number: 6304431, assigned to Samsung Electronics, Oct 01). The invention makes the keyboard more ergonomic without surrendering the portability.

According to the invention, the keyboard unit is attached to an upper surface of the main body and is structurally and operationally separated along a central axis into a left keyboard half and a right keyboard half. Both the keyboard halves are coupled to the main body but rotate on the first and second hinges to be suitably positioned according to user requirement.

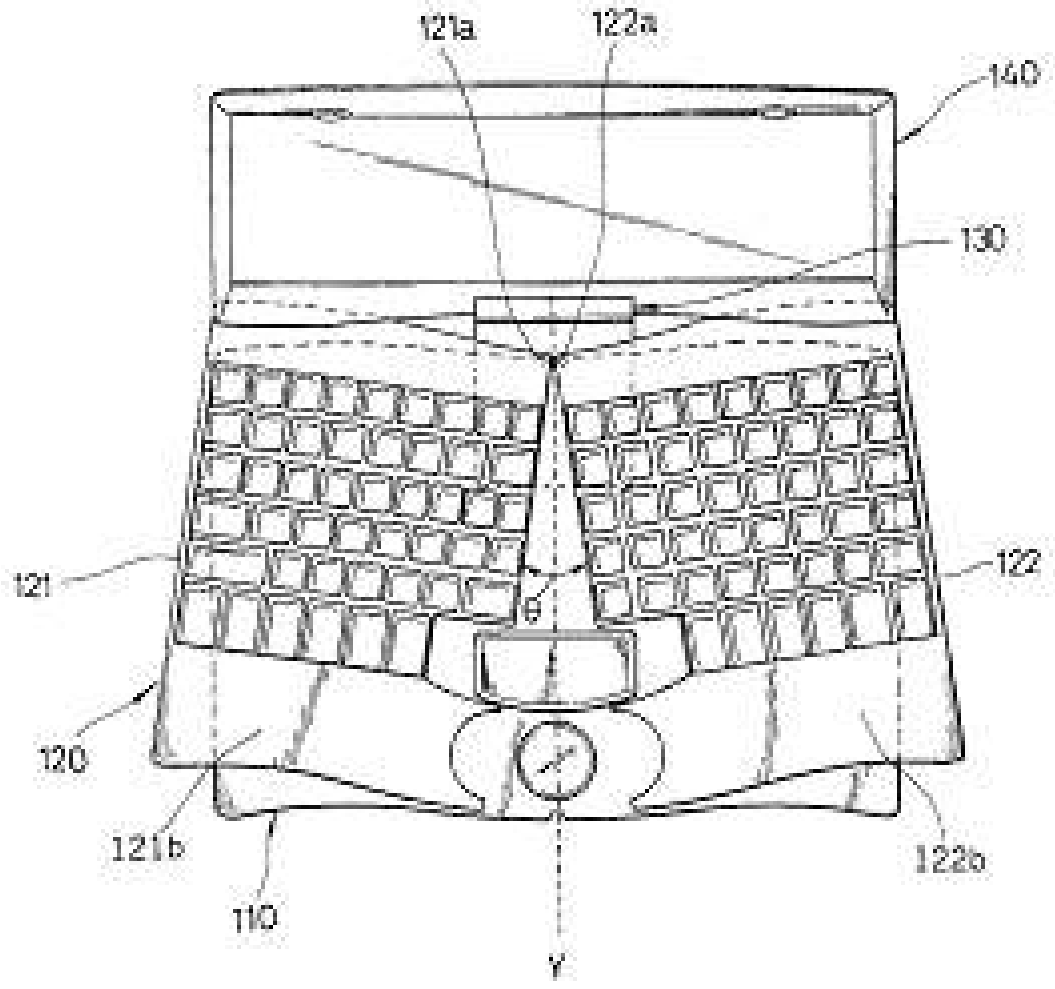

## TRIZ based analysis

The invention splits the keyboard into two, one for each arm **(Principle-1: Segmentation)**, and make them flexible to adjust with the angle of the arm. **(Principle-15: Dynamize)**.

## 2.10 Portable electronic device with a concealable keyboard module (Patent 6556430)

## Background problem

The conventional model of notebook computers includes a keyboard and a touch pad, which covers all the space on the surface and leaves no space to install other controls. It is necessary to create space on the laptop surface to install other controls like touch pad, pen control etc.

## Solution provided by the invention

Kuo et al. invented a concealed keyboard (Patent Number: 6556430, assigned to Compal electronics, Apr 03) which can be used with a notebook or PDA or similar small devices. The keyboard remains inside the main module and can be pulled out via an opening. This mechanism leaves enough space for the touch control, pen control and other devices to be fixed on the main surface of the laptop.

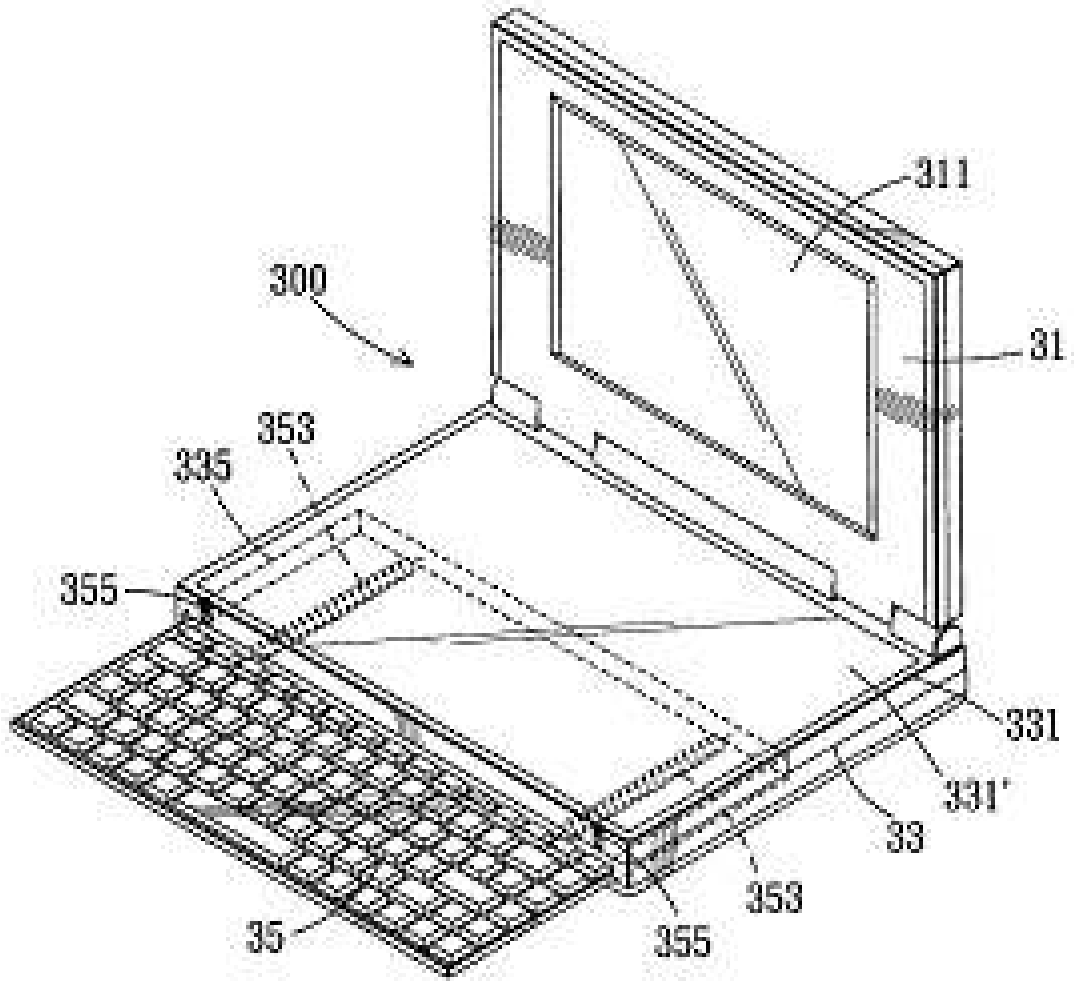

**TRIZ based analysis**

The invention makes a drawer type keyboard concealed within the laptop, which can be pulled or pushed easily according to usage **(Principle-34: Discard and Recover)**.

The surface of the keyboard is used for installing other controlling devices **(Principle- Another Dimension)**.